\documentstyle{article}
\date{}             
\begin{document}
\title{The $\omega NN$ couplings derived from QCD sum rules}
\author{Shi-Lin Zhu\\
Institute of Theoretical Physics, Academia Sinica\\ 
P.O.BOX 2735, Beijing 100080, China\\
FAX: 086-10-62562587\\
TEL: 086-10-62541816\\
E-MAIL: zhusl@itp.ac.cn}
\maketitle
\begin{center}
\begin{minipage}{120mm}
\vskip 0.6in
\begin{center}{\bf Abstract}\end{center}
{\large
The light cone QCD sum rules are derived for 
the $\omega NN$ vector and tensor couplings 
simultaneously. The vacuum gluon field contribution is taken into account.
Our results are $g_\omega =18\pm 8$, $\kappa_\omega =0.8\pm 0.4$.

PACS Indices: 21.30.+y; 13.75.Cs; 12.38.Lg


}
\end{minipage}
\end{center}

\large

Recently the $\rho NN$ couplings were calculated in the framework of QCD sum 
rules using vector meson light cone wave functions \cite{zhu-nucl}. 
Due to isospin symmetry the contribution from the vacuum gluon fields, which 
appears as the three particle rho meson wave functions, cancels exactly in the
sum rules for the $\rho NN$ couplings. In this note we extend the same 
formalism to extract $\omega NN$ couplings in QCD. One big difference is that 
the vacuum gluon fields play an important role in the present case.

We omit the detailed derivation and present the final light cone sum rules 
and numerical results for $\omega NN$ couplings directly.
Denote the $\omega NN$ vector coupling by $g_\omega$ and the tensor-vector
coupling ratio by $\kappa_\omega$. We have,
\begin{eqnarray}\label{lon-3}\nonumber
\lambda^2_N \sqrt{2} g_\omega {1+\kappa_\omega\over 2}m_\omega^2
e^{-{ m_N^2\over M^2} } = {1\over 2\pi^2}
e^{-{u_0(1-u_0)m_\omega^2\over M^2}} \{
-m_\omega g^v (u_0) M^6 f_2 ({s_0\over M^2})
&\\ \nonumber
-{1 \over 3} f_\omega m_\omega
\phi_\parallel (u_0)  M^6 f_2 ({s_0\over M^2})  
+3f_\omega m_\omega^3 G_3 (u_0) M^4 f_1 ({s_0\over M^2})  
&\\ \nonumber
+f_\omega  m_\omega^3 A (u_0) M^4 f_1 ({s_0\over M^2}) 
-{1\over 24} f_\omega m_\omega \langle g_s^2G^2 \rangle
g^v (u_0) M^2 f_0 ({s_0\over M^2})
&\\ \nonumber
+{ 1\over 2} f^V_{3\omega}
[{1\over 2}I_0^v M^6f_2 ({s_0\over M^2}) 
+m_\omega^2 I_2^v M^4f_1 ({s_0\over M^2})]
&\\ 
+ { 1\over 2} f^A_{3\omega}
[-{1\over 2}I_0^a M^6f_2 ({s_0\over M^2}) 
+m_\omega^2 I_1^a M^4f_1 ({s_0\over M^2})]
\}&\; ,
\end{eqnarray}
\begin{eqnarray}\label{lon-4}\nonumber
\lambda^2_N \sqrt{2} g_\omega 
e^{-{ m_N^2\over M^2} } = {1\over 2\pi^2}
e^{-{u_0(1-u_0)m_\omega^2\over M^2}} m_\omega \{
-{2 \over 3} f_\omega \phi_\parallel (u_0) M^4 f_1 ({s_0\over M^2})
&\\  \nonumber
+6f_\omega  m^2_\omega G_3(u_0) M^2 f_0 ({s_0\over M^2})
+2f_\omega  m^2_\omega A (u_0) M^2 f_0 ({s_0\over M^2})
&\\ 
+ f^V_{3\omega}m_\omega I_2^v M^2f_0 ({s_0\over M^2})
+f^A_{3\omega}m_\omega I_1^a M^2f_0 ({s_0\over M^2})
\}&\; ,
\end{eqnarray}
where $f_n(x)=1-e^{-x}\sum\limits_{k=0}^{n}{x^k\over k!}$ is the factor used 
to subtract the continuum, $s_0$ is the continuum threshold,
$\phi_\omega^\prime (u_0)  ={d\phi_\omega (u)\over du}|_{u=u_0}$ etc.
Since the initial and final states are the same, the sum rules are 
symmetric with the Borel parameters $M_1^2$ and $M_2^2$. 
It's reasonable to adopt $M_1^2=M_2^2=2M^2$, i.e., $u_0 ={1\over 2}$. 
Such a symmetric choice enables a clean subtraction of the continuum and 
excited states contribution and leads to the above relatively 
simple expressions. We shall work in the physical limit $q^2=m_\omega^2 $.

We have defined
\begin{equation}\label{ggg}
G_3 (u)=\int_0^u dt \int_0^t ds C (s) \; ,
\end{equation}
The functions $I_i (u_0), i=0, 1, 2, 3, 4$ are:
\begin{eqnarray}\label{iv0}\nonumber
I_0^v =-2 \int {\cal D}\underline{\alpha} {{\cal V}(\alpha_i )\over \alpha_g^2}
[\delta(\alpha_g +\alpha_3-u_0) -\delta(\alpha_3 -u_0)] 
& \\ \nonumber
+\int_0^1d\alpha_g \int_0^{1-\alpha_g} d\alpha_3 {1\over \alpha_g}
{d\over d\alpha_3}{\cal V}(1-\alpha_3 -\alpha_g) 
[\delta(\alpha_g +\alpha_3-u_0) -\delta(\alpha_3 -u_0)] 
& \\ 
+\int_0^1 d\alpha_g {{\cal V}(0,\alpha_g, 1-\alpha_g ) \over \alpha_g}
\delta (1-u_0-\alpha_g) 
-\int_0^1d \alpha_g {{\cal V}(1-\alpha_g,\alpha_g, 0 ) \over \alpha_g}
\delta (\alpha_g-u_0)
&\; ,
\end{eqnarray}
\begin{equation}\label{ia0}
I_0^a=+\int_0^1 d\alpha_g {{\cal A}(0,\alpha_g, 1-\alpha_g ) \over \alpha_g}
\delta (1-u_0-\alpha_g) 
+\int_0^1d \alpha_g {{\cal A}(1-\alpha_g,\alpha_g, 0 ) \over \alpha_g}
\delta (\alpha_g-u_0)
\; ,
\end{equation}
\begin{equation}\label{i1}
I_1^F=\int_0^1du \int {\cal D}\underline{\alpha}
{\cal F}(\alpha_i) \delta (u \alpha_g +\alpha_3 -u_0)
\; ,
\end{equation}
\begin{equation}\label{i2}
I_2^F=\int_0^1du \int {\cal D}\underline{\alpha} (1-2u)
{\cal F}(\alpha_i) \delta (u \alpha_g +\alpha_3 -u_0)
\; ,
\end{equation}
where ${\cal F}={\cal V}, {\cal A}$ respectively. The definitions 
of the other vector meson wave functions (VMWFs) can be found in \cite{braun98}.

At $u_0 ={1\over 2}$ we have \cite{braun98}
$g^v=0.64$, $\phi_\parallel (u_0)=1.1$,
$G_3 (u_0)=-0.13$, $A (u_0)=2.18$,
$I_0^v=262.5$, $I_1^a=2.04$, $I_2^v=0.4375$,
$I_0^a=0$, $I_1^v=0$, $I_2^a=0$ at the scale $\mu =1 $GeV.

Numerically we have:
\begin{equation}\label{num-1}
g_\omega (1+\kappa_\omega )= (45 \pm 9)\; ,
\end{equation}
\begin{equation}\label{num-2}
g_\omega = ( 26\pm 6)\; .
\end{equation}

Dividing (\ref{lon-3}) by (\ref{lon-4}) we get a new stable 
sum rule for $1+\kappa_\omega$. 
\begin{equation}\label{num-3}
1+\kappa_\omega =( 1.7\pm 0.4)\; ,
\end{equation}
which corresponds to
\begin{equation}\label{num-33}
\kappa_\omega =( 0.7\pm 0.4)\; .
\end{equation}

The major uncertainty comes from the VMWFs since our final sum rules depend 
both on the value of WFs and their integrals at $u_0$. Especially,
the sum rules (\ref{lon-3}) and (\ref{lon-4}) are sensitive to the 
variations of the values of VMWFs  at $u_0={1\over 2}$.
For example, we will get $g_\omega = (10 \pm 2)$ if we use the asymptotic 
form for the VMWFs. However, the ratio of these
two sum rules is insensitive to the specific form of these VMWFs,
$\kappa =1.0\pm 0.4$ for the asymptotic form of VMWFs. If we take  
into account the uncertainty in VMWFs by treating the asymptotic form and
the model wave functions as two opposite limits for the real VMWFs,
we obtain:
\begin{equation}
g_\omega = ( 18\pm 8)\; ,
\end{equation}
\begin{equation}
\kappa_\omega =( 0.8\pm 0.4)\; .
\end{equation}

These results agree with a recent dispersion-theoretical 
analysis of the nucleon electromagnetic form factors \cite{mergell},
where vector meson nucleon couplings were extracted rather precisely:
\begin{equation}
g_\omega = ( 20.86 \pm 0.25)\; ,
\end{equation}
\begin{equation}
\kappa_\omega =( -0.16\pm 0.01)\; .
\end{equation}

It is interesting to notice that vacuum gluon fields play an 
important role in the vector meson nucleon interaction. Our result 
supports the large value for $\kappa_\rho$, $g_\omega$.
The naive relation $g_\omega =3g_\rho$ does not hold any more. 
We want to emphasize that there are no free parameters in our calculation
once the values of the VMWFs at the point $u_0={1\over 2}$ are 
determined, which is constrained by the QCD sum rule analysis of 
their moments to some extent \cite{braun98}.

\vspace{0.8cm} {\it Acknowledgments:\/} This project was 
supported by the Natural Science Foundation of China.


\bigskip
\vspace{1.cm}

\begin{thebibliography}{99}
\bibitem{zhu-nucl}Shi-Lin Zhu, Phys. Rev. C 59, 435 (1999).

\bibitem{mergell}P. Mergell, U.-G. Meissner, and D. Drechsel, 
Nucl. Phys. A. {\bf 596}, 367 (1996).

\bibitem{braun98}P. Ball et al., Nucl. Phys. B 529, 323 (1998); 
P. Ball and V. M. Braun, hep-ph/9808229.

\end{thebibliography}
\end{document}